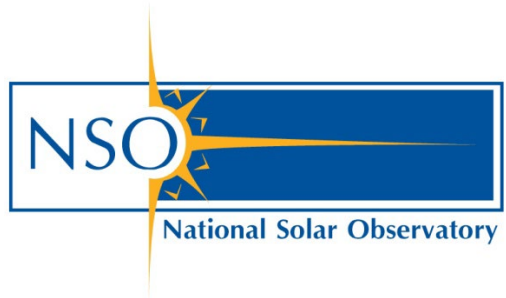

# GONG third generation camera:
# Detector selection and feasibility study

Sanjay Gosain, Jack Harvey, Detrick Branson, Steve Bounds, Tim Purdy, Sang Nguyen, Greg Card

Institution(s)

National Solar Observatory, Boulder, Colorado

Date (October-13-2022)

Technical Report No. **NSO/NISP-2022-003**

**Abstract:** Aging GONG second generation cameras (Silicon Mountain Design™ cameras) were planned to be replaced after their long service of more than a decade. This prompted a market-wide search for a potential replacement detector to meet the GONG science requirements. This report provides some history of the search process, a comparison between CMOS and CCD type sensors and then a quantitative evaluation of potential candidates to arrive at final selection. Further, a feasibility study of the selected sensor for adaptation to GONG optical system was done and sensor characteristics were independently verified in the laboratory. This technical report gives description of these studies and tests.

**Introduction:** GONG SMD cameras were deployed in service around 2005. These cameras were based on CCD (charge coupled device) technology, more specifically, they employed four channel frame transfer CCD. The camera was designed by Silicon Mountain Design (SMD), a company which was later acquired by Dalsa™. These cameras were uniquely qualified to meet GONG detector requirements due to a combination of high frame rate (60fps) and high full well (>200ke-) of the pixels (*Harvey 1996; Harvey et al. 1998*). Such high-speed commercial cameras are typically designed for machine vision applications for the manufacturing industry to enable, for example, inspection of products in the assembly line.

Over last few decades, a new imaging sensor technology called CMOS (complementary metal oxide semiconductor) sensors, has become quite widespread and are becoming comparable to CCDs in terms of optical performance. Due to their widespread use in commercial sector, i.e., in smart phone devices and low-light digital microscopy applications, their cost is much less compared to CCD due to mass production. Therefore, it is no surprise that most commercial camera manufacturers, whose biggest consumer base is the machine vision and biomedical imaging industry, have transitioned in recent years to CMOS sensors. Due to GONG budget constraints a custom camera development around a CCD sensor was deemed too expensive. Fortunately, we were able to find a good CCD sensor replacement from Dalsa™ along with board level camera electronics, however, by the time we could evaluate the sensor in a laboratory, the company obsoleted the product and moved on to CMOS technology. This meant NSO had no other option but to consider CMOS based commercial-off-the-shelf (COTS) cameras for GONG.

Here we briefly discuss the difference between CCD and CMOS sensors. In both sensors, the so-called "photosites" convert incident photons into electrons. This accumulated electronic charge needs to be read out as electrical signal. This is where key differences arise. In a CCD the charge is transported across the chip and read (converted to analog voltage) at one corner of the pixel array and digitized externally to the sensor using Analog to Digital Convertors (ADCs). In most CMOS devices there are several transistors at each photosite that convert charge to voltage at each pixel. The signal is then multiplexed by row and column to multiple on-chip ADCs. This makes the sensor more flexible as each photosite can be read individually. Therefore, CMOS sensor is sometimes also called Active Pixel Sensor (APS). CCD sensors use special design in their manufacture that allows them to transfer charge undistorted leading to high image quality and high sensitivity. However, there has been significant advances in CMOS technology over the years that has narrowed the gap with CCD in terms of image quality and sensitivity.

In the following sections we describe the detector requirements for GONG, selection criterion for COTS camera, market survey, detector selection and feasibility study for GONG optical system.

**Detector requirements:**

The Table 1 lists the detector requirements for GONG.

| Sensor format | At least 1024x1024 | To maintain data format consistency |
|---|---|---|
| Pixel full well | ~100 ke- or better | To utilize high photon flux in GONG optical system |
| Frame rate | ~60 or higher fps | Same as above |
| A/D conversion | 12 bits or higher | To avoid quantization noise. |
| Read noise | ~ few tens of electrons | To maintain useful dynamic range for distinguishing small changes in intensity. |
| Pixel linearity error | < 1% | To avoid spurious phase errors from pixel gain nonlinearity. |
| Optical fill factor | ~1 physically or optically via micro lens | To avoid aliasing of higher frequencies. |
| Physical pixel size | More than 7 microns | To maintain reasonable physical distance between camera rotator and waveplate rotator assembly. |

**Selection criterion other than detector specifications:**

In addition to the detector specifications mentioned above we also required following constraints:
1. Cost of camera and accessories: Below $10,000 per camera.
2. Temperature sensor to monitor CMOS temp: Required
3. Length of data cable between camera and PC: ≥ 5 meters
4. Software for data acquisition: Software development kit for custom camera environment development.
5. External trigger for exposure synchronization and sync out:
6. Global electronic shutter
7. Operating temperature of less than 60 deg C (via active or passive cooling).

8. Thermal stability of better than 5 deg C during continuous operation.
9. Exposure time jitter of less than few microseconds.
10. Technical support and product warranty.
11. Reasonable delivery schedule for a minimum of 10 cameras.
12. Spare parts such as sensor boards and interface cards.
13. Compatibility with Linux (Fedora OS).

**Market survey and shortlisting of potential candidates:**

Initially a thorough survey of various cameras and sensors was carried out. Full details of each of these candidates is not of much interest, however, key performance benchmark of SNR is summarized in the bar chart shown in Figure 1. The name of candidate camera/sensor is labeled in the abscissa and the SNR achieved over a total integration time of 10 seconds, which is equivalent integration time for each of the six intensity buffers, is shown on the ordinate. The SNR requirement of 6000 for GONG is marked with a horizontal solid line. About ten of these candidates were found to meet the SNR requirement.

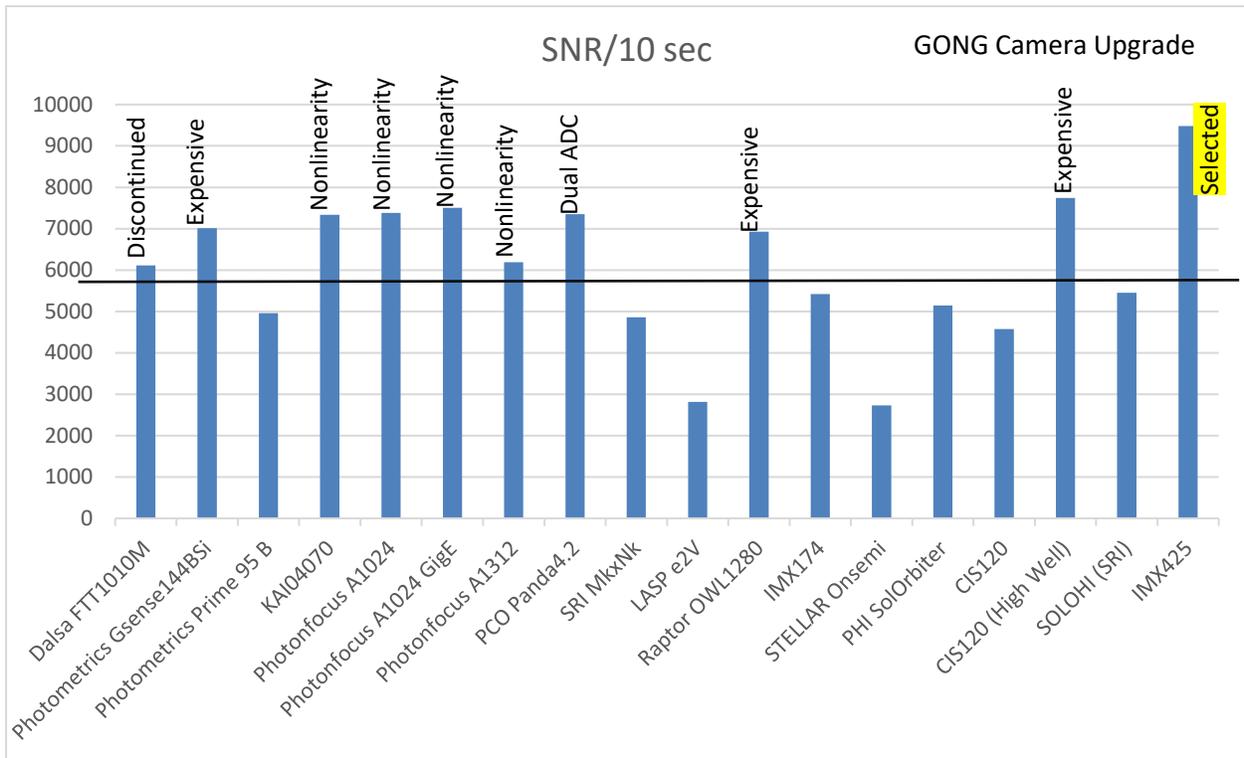

*Figure 1: Comparative SNR of various sensor/camera and the reason for not choosing potential candidates.*

Sony IMX425 provided the best performance with significant margin above our requirement. Initially, only one vendor was developing cameras using this sensor, however, multiple vendors started incorporating this sensor in their product line while we did our trade study.

Fully developed cameras rather than sensors, that met the specifications, were given preference. Based on the SNR criterion (see Figure 1) a handful of cameras were rejected merely due to being too expensive, as indicated in Figure 1. We pursued following cameras eventually, where we studied them in more detail and had several technical exchanges with the vendors before finalizing the fourth option below.

1. Photonfocus™ : Model -  MV1-D1024E-160-CL
2. PCO™ : Model - Edge 5.5 camlink HS
3. Matrix-vision™ : Model – mvBlueCougar-X102mG
4. Emergent Vision Tech. ™ : Model EVT HS1800

Out of these candidates the first camera from PhotonFocus™ had good characteristics on paper, however, the two main issues with it were a) nonlinear pixel response of more than 2% and, b) a fill factor of 30%. Also, it had a fixed pattern noise that was too high.
The PCO camera had a good linearity and noise characteristics, however, this sensor used dual ADC, a high and a low gain channel with 11 bits each, and later combined output to create 16-bit output. This created an issued with a slight bump for signals near the middle of the dynamic range.

It was clear from the beginning that Sony IMX 425 sensor, which was released during the beginning phase our study, had promising specifications, and exceeded our requirements. However, COTS cameras based on this sensor were not available initially. However, soon two COTS cameras were announced (item 3 and 4 in the list above). While Matrix-Vision camera has the same sensor as EVT camera, the 25 GigE data interface provided with EVT (as compared to only 1 GigE interface with Matrix-Vision camera) allows to unlock the full potential of this sensor and was, therefore, finalized for GONG upgrade. Further, a fiber optic interface provided with EVT cameras was deemed attractive as it allowed for longer cable lengths (up to 100m) and a RFI free data transmission path.

**Laboratory Characterization of EVT camera with IMX425 sensor:**
The sensor characteristics published by the camera vendor were independently verified via laboratory characterization. Since, each batch of sensors can vary in properties we needed to test the characteristics for our batch of sensors and to spot any possible odd ones. Also, such tests can spot any odd effects introduced by the readout electronics and any camera firmware issues.

The following tests were conducted:
1. Photon transfer curve measurement to derive readout noise, gain, linearity etc.
2. Estimate noise floor, SNR and dynamic range.
3. Photon response non uniformity

4. Dark signal non-uniformity.

**Test setup at Laboratory:** The sensor characteristics are wavelength dependent, so we used a LED emitting near GONG operating wavelength of 670 nm. Further, to stabilize the output we used a regulated power supply and passive thermal stabilization. Specifically, we used thermally stabilized DC powered LED from Thorlabs™. The schematic of the setup is shown in Figure 2. below.

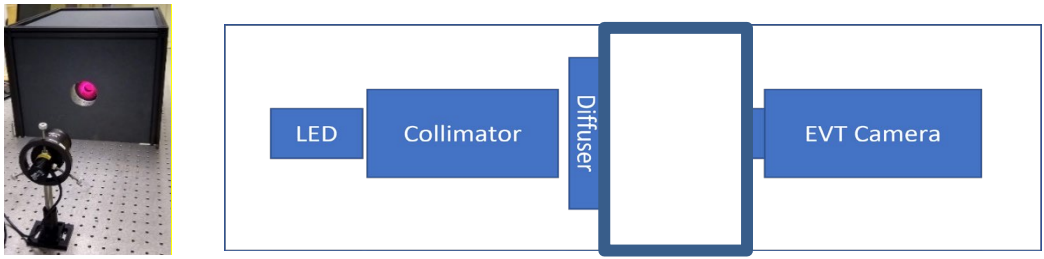

*Figure 2. Right panel shows he schematic of the test setup for sensor characterization. Note how the diffuser and camera are mounted on the ports at the opposite ends of a light-tight box. Left panel shows the stabilized LED source and collimator assembly on a optical bench postholder and the illuminated (reddish) diffuser mounted on a port of the dark box. The camera is not seen in the picture and is on the opposite end of the box.*

A collimator is attached to the LED to bring the diverging optical beam into collimation. This collimated beam then falls on an optical diffuser glass plate. This diffuser plate is mounted on an entrance port of a dark box, while the camera to be tested is mounted on the other port, at the opposite end of the box. Camera is kept at a distance from the diffuser such that it sees a spatially uniform illumination. Finally, the temporal stability of the light source is also very important. With our setup LED output was measured to be stable temporally for 30 seconds up to 0.01% RMS. This spatial and temporal stability adequately addresses the basic requirements of the sensor characterization tests. The Figure 3 shows the spatio-temporal stability of the light source.

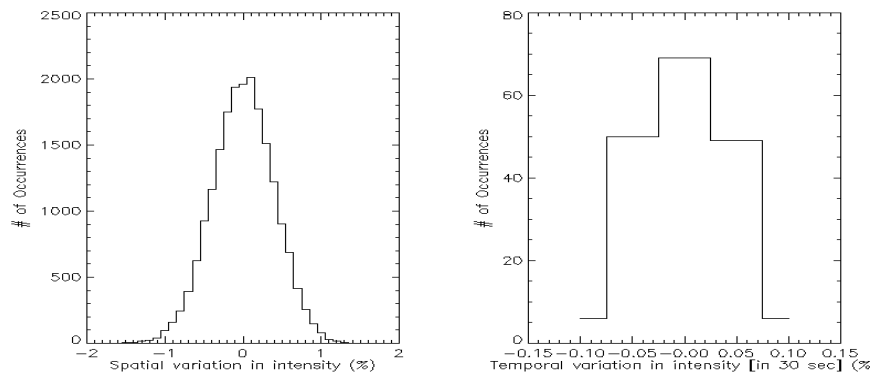

*Figure 3 : Left panel shows a histogram of the spatial variation of the light intensity, and the right panel shows temporal variation during 30 seconds (time to acquire test data).*

**Camera sensor tests:** Laboratory measurements were made to derive readout noise, gain factor, camera offset, nonlinearity of pixel response over the full dynamic range etc. The measurements consist of keeping the illuminating light source steady while varying the exposure time from zero to near saturation level in equal steps. The camera was set up in a mode where the data format was monochrome 12 bit output these images were acquired with a lowest gain setting and a DC offset setting of zero.

*Photon Transfer Curve:*

Using these measurements, we derive the photon transfer curve (PTC). The advantage of PTC is that absolute measurement of light input is not required (Janesick, 2007). In Figure 4 below, we plot the PTC, which represents temporal noise of the pixels as a function of the mean signal in log-log plot. Typically, this plot has three parts which are marked by vertical dashed lines in the Figure 4, (i) a read noise dominated part on the left corresponds to low signal levels, (ii) photon shot noise dominated part for higher signal levels (represented by a straight line fit of slope 0.5), and (iii) part where signal causes pixel to saturate or exceed full well capacity of the pixel.

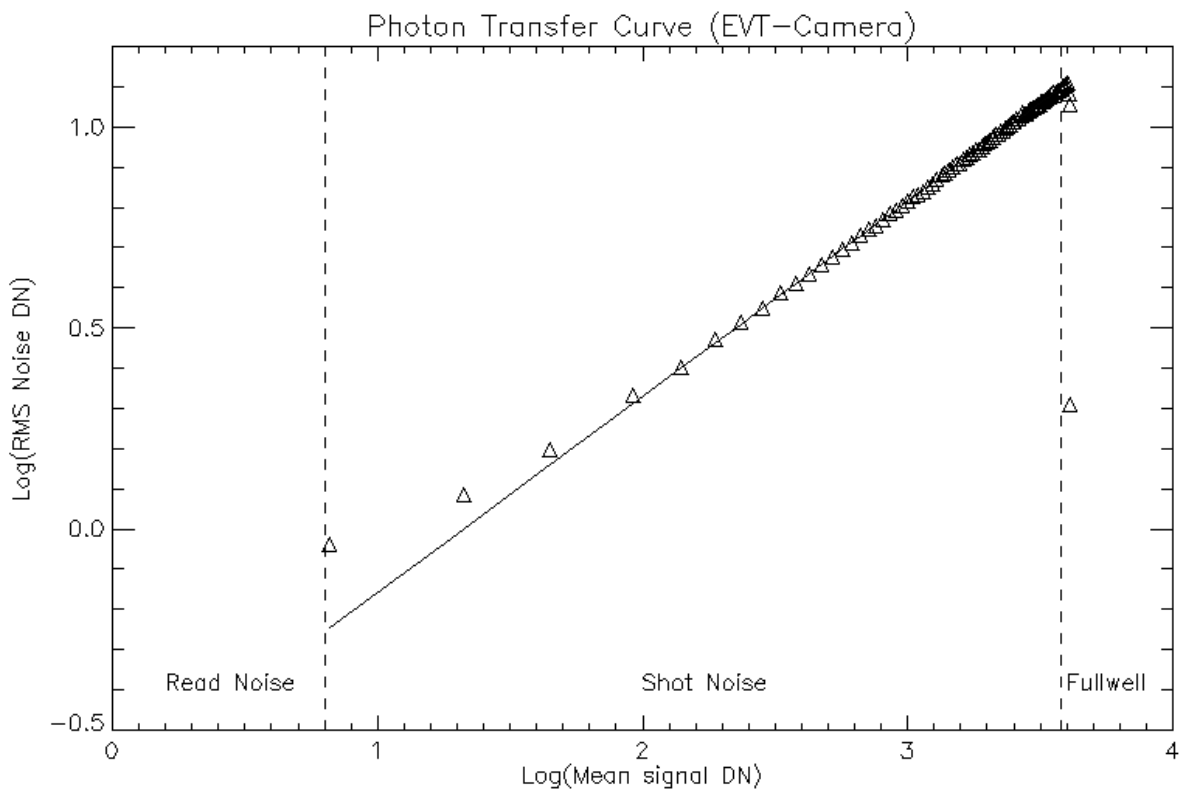

***Figure 4.:*** *Photon transfer curve is shown which is standard deviation versus mean signal.*

*Read Noise:*

Using this curve, we can determine the read noise of the detector which is the flat part of the curve in the left part of the plot. Since photon signal is very low in this part, the noise is dominated by the read noise of the electronics, this is sometimes also known as temporal dark noise in the literature.

*Pixel gain (e-/ADU):*

The PTC observations also provide a way of determining gain of the pixels, that is, number of photoelectrons per data number. To obtain gain, we note that for an increase in illumination of X, the camera average signal level will change by GX, while the noise variance changes by $G^2X$. The ratio of these two is the slope of the mean signal versus variance curve and provides the pixel gain conversion factor. Figure 5 below shows an example of measured gain curve for our camera detector.

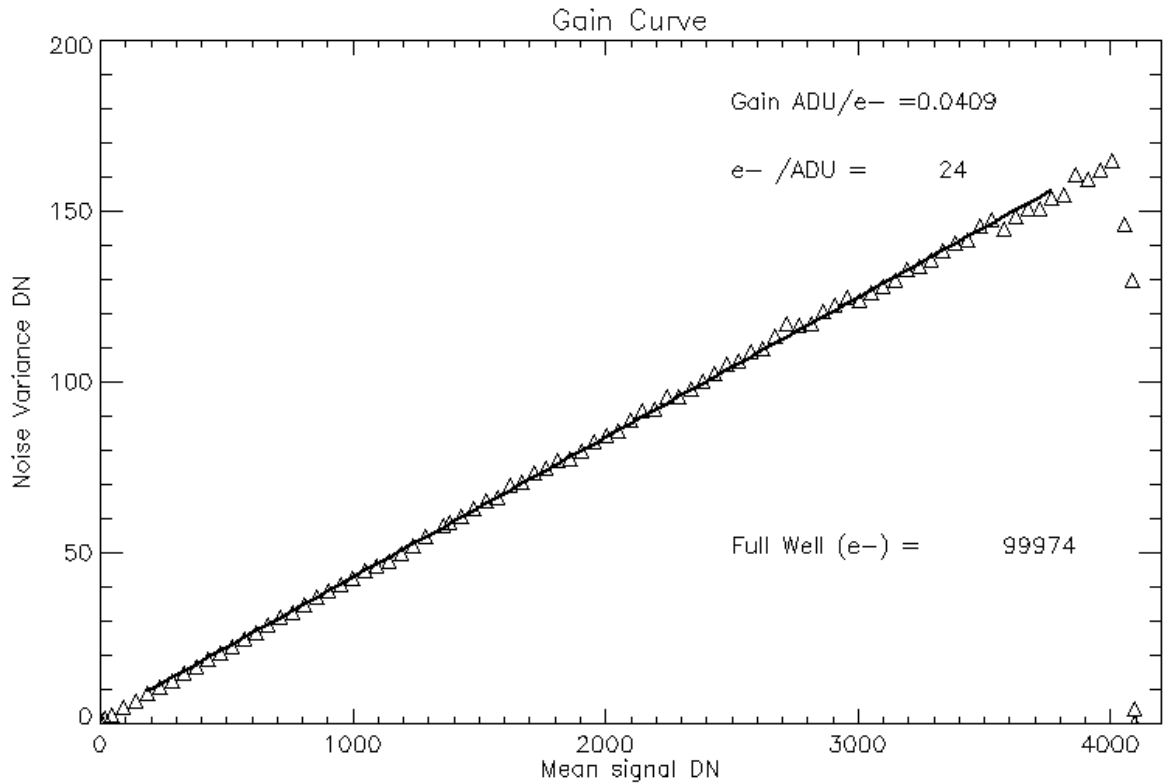

*Figure 5:* *Gain conversion factor is derived from the slope of the mean signal versus variance curve.*

*Pixel Full well:* Once the gain conversion factor is known we can estimate the pixel full well by multiplying maximum ADC digital count and gain factor. For the example curve in Figure 5 the gain factor G=24e-/ADU and maximum digital depth of 12 bit gives an estimate of a full well of about 100ke-.

*Pixel nonlinearity:* The mean digital output of an ideal digital sensor should increase linearly with increasing illumination level (in our case this means linearly proportional to the exposure time as the light source is kept stable). In real detectors the response is, however, not perfectly linear. A linear fit to the measured response curve provides a means of determining departures from linearity. This is known as integral nonlinearity (to distinguish it from differential nonlinearity which may arise due to bit error, i.e., non-uniform ADC step size). This integral nonlinearity is defined as follows: INL=100*($E_{max}$-$E_{min}$)/$AD_{FS}$ Where $E_{max}$ and $E_{min}$ are maximum and minimum departures from the best-fit straight line and $AD_{FS}$ is AD converter full-scale value (counts). Figure 6 shows the linearity response curve from our camera tests. The INL is found to meet our requirements of better (smaller) than 1%.

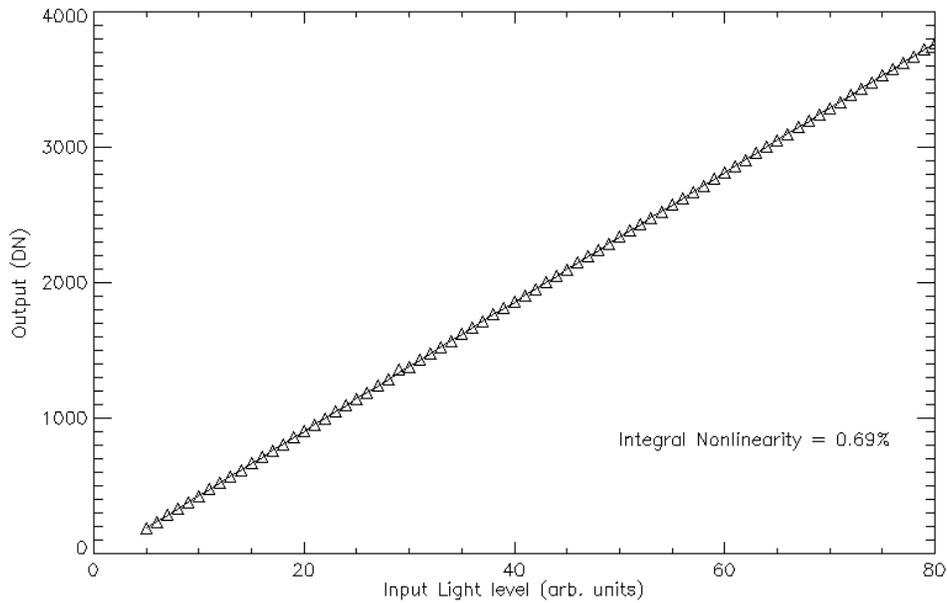

*Figure 6:* *Linearity response curve of the detector and best-fit straight line.*

*Dark and Flat Response:* The top panels in Figure 7 below shows the mean flat and dark maps after averaging over 100 frames with a millisecond exposure. The flat field map is shown in reverse colormap. The spatial distribution of gain and dark signal non-uniformity is shown in the bottom panel histograms. The gain values are normalized to median value of the flat field while dark signal is shown as is. The gain and dark signal is quite uniform spatially.

*Hot and Dead pixels:* In addition, we also determined dead and hot pixel statistics and found only 0.02% dead pixels (defines as 4-sigma below median value of 1) and 0.005% hot pixels (4-sigma above median value of 1) in the gain distribution, in the entire pixel array 1.76 million pixels.

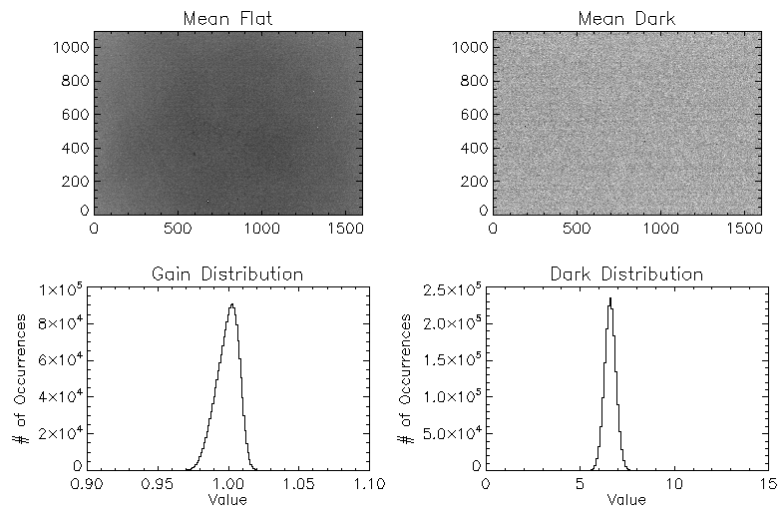

*Figure 7:* *Top panels show spatial distribution maps of flat-field and dark signal over entire pixel array. Bottom panels show the histogram of the gain distribution and dark values.*

**SNR performance estimate:**

Based on the measured sensor characteristics we estimate that the EVT camera can provide a higher SNR in a one-minute integration due to a high pixel full well and frame rate (we assume 180 fps here, however, the sensor can achieve up to 240 fps). These tests also show a linear behavior and adequate full well and frame rate to meet the SNR requirements.

| Current SMD Camera | New EVT Camera |
|---|---|
| 200 ke- | 100 ke- |
| 60 fps | 180 fps |
| 12 bit | 12 bit |
| SNR/pixel/intensity buffer ~8,500 | SNR/pixel/intensity buffer ~10,000 |

**Optical integration feasibility:**

The optical integration of the new cameras to current GONG system required adaptation of image plate scale via new reimaging lenses. This was done using the Zemax™ optical design software. While doing this the following things were kept in mind:

(i) The distortion across the full circular field-of-view (FOV) of radius 0.27 degrees should be kept to a minimum (better than 0.1%).
(ii) There should be no optical vignetting due to camera C-mount adaptor.
(iii) The spot diagram should show diffraction limited performance across the FOV.
(iv) The MTF and PSF should not change from the current design.
(v) Provide tolerances to mechanical engineer about depth of focus, misalignment in tip-tilt of camera and decentration.

*Lens Specifications to adapt Solar Image to new Camera format:*

Using Zemax™ optical design software it was found that the task could be accomplished with a stock lens from OptoSigma SLB-25.4-80PM with standard visible anti-reflection coating with following specifications: 25.0mm Dia. x 80.0mm Focal Length, VIS 0° Coated, Plano-Convex Lens.

| Parameter | OLD | NEW |
|---|---|---|
| Pixel size | 14 micron | 9 micron |
| Working f/D | 17 | ~11 (10.86) |
| Depth of focus (Normalized to R_sun) | 476 micron (0.0867) | 196 micron (0.056) |
| Tolerance on tilt of sensor due to depth of focus | 3.79 deg | 2.48 deg |
| Plate Scale | ~2.5" per pixel | ~2.5" per pixel |
| Solar Diameter | 786 pixels | 778 pixels |

*Tolerancing of lens alignment for mechanical design:*

- Tolerance in decenter of the lens is about +/- 300 microns in order to keep the image motion within 10 pixels (if more than 10 pixels is tolerable then it can be more).
- Distortion and spot diagram are quite robust with lens decentration (up to +/- 2 mm).
- Tip-Tilt tolerance of the lens is +/- 0.6 degrees for distortion to stay within 0.1%.
- Spot diagram is robust up to tip-tilt of +/- 4 degrees.

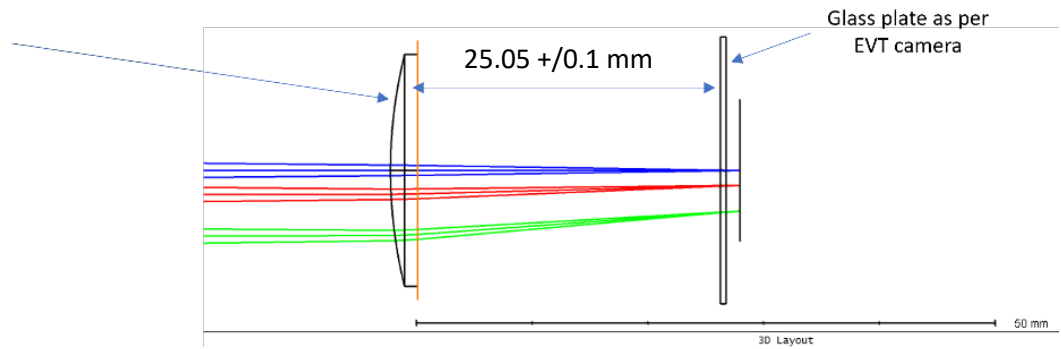

**Figure 8:** *The ray trace of stock lens placed just in front of the camera (potentially mountable in the C-mount collar with adaptor lens tubes).*

*Field curvature and Geometric Distortion:*

The resulting optical distortion with the stock lens is expected to be around 0.02% near the limb and even smaller near the center of the solar disk as shown in the right panel of Figure 9 below. This is well below the requirement of 0.1% over the full field of view (FOV). The field curvature ranges about 100 microns from center to edge of FOV. The Figure 10 below shows that the spot diagram is diffraction limited within a depth of focus of +/- 100 microns.

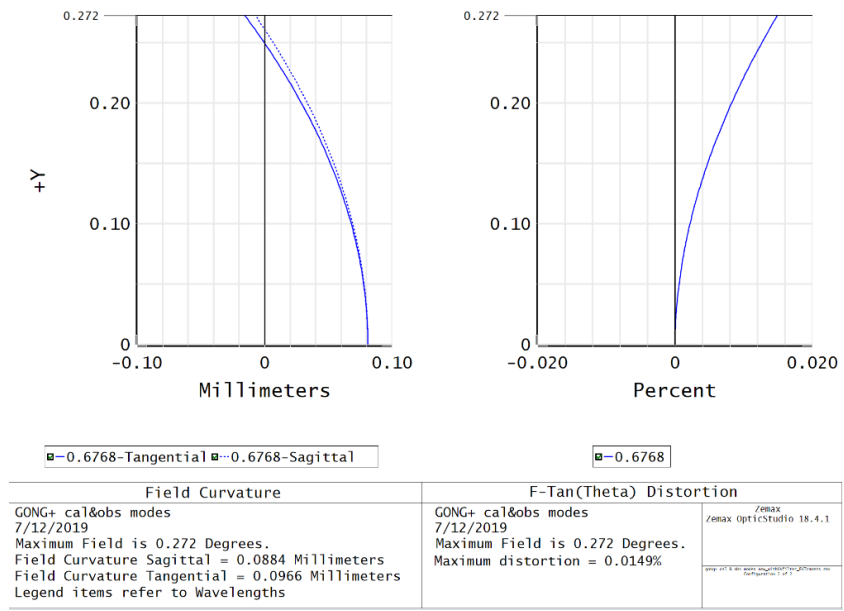

**Figure 9:** *Field curvature and distortion as a function of center to edge of field-of-view.*

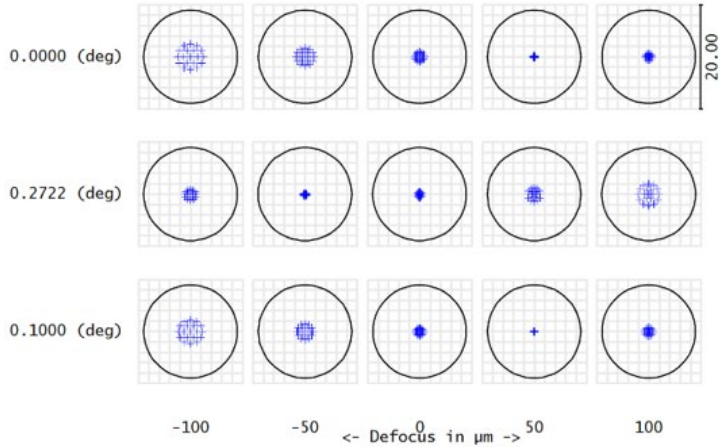

*Figure 10: The spot diagram with depth of focus in a range of +/- 100 microns.*

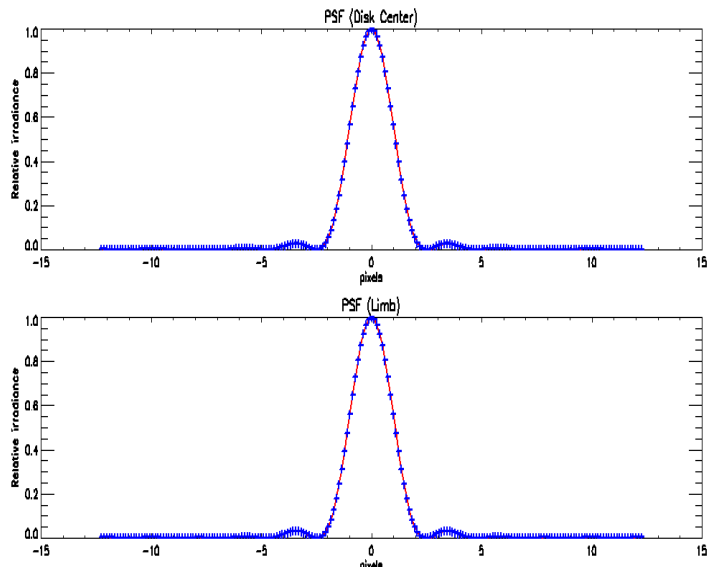

*Point Spread Function:* PSF of old and new system is compared to study the impact of new lens modifications on the optical performance. Figure 11 below shows the PSF of the old and new optical system. X-axis is respective pixel number. Y-axis is relative irradiance. Red curve and blue symbols represent the trace of PSF for the old and new optical design, respectively. It shows that the PSF performance is maintained with the new lens modifications.

*Figure 11: PSF of the old (red curve) and new system (blue symbols) at the center (top panel) and edge of the FOV (bottom panel) are shown.*

**Thermal Tests:**

EVT camera does not have active cooling mechanism (heat sinks are provided). An external sensor mounted on the outer surface of camera body showed that after powering ON the camera, starting at a room temperature of 23.8 deg. C, the camera reaches thermal equilibrium with ambient of about 46 deg. C in about 90 minutes, regardless of the frame rate. We tried the same test with 60 and 180 frames per second and found similar performance. The stability of temperature to within a couple of degrees after reaching equilibrium means we can expect the dark current to be stable for long observing sequences.

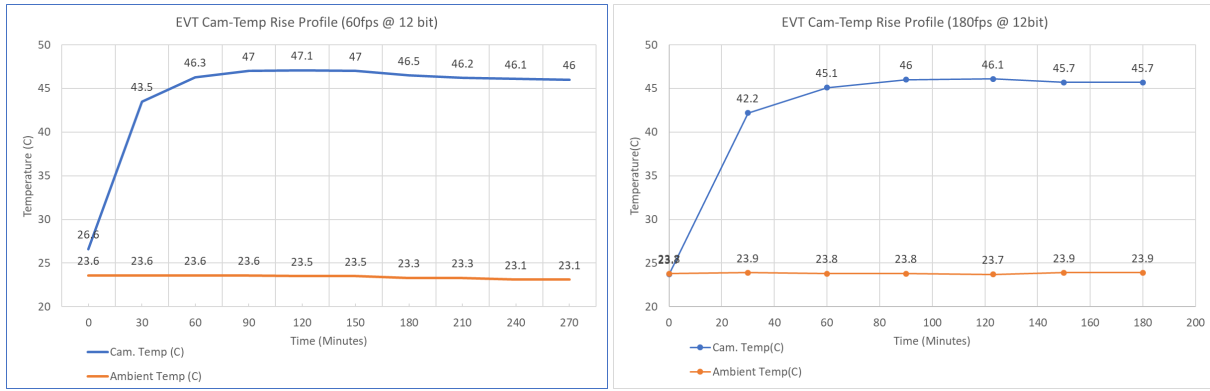

*Figure 12: Temperature of camera body after turning ON from overnight OFF position. The temperature curve on left and right correspond to different frame rates.*

**Summary:**

This study summarizes the results of sensor characterization tests and optical design modifications study regarding the feasibility of the selected sensor/camera as a potential replacement for GONG SMD cameras. The CMOS sensor demonstrates excellent linearity and spatial uniformity in its signal response. The dark uniformity and read noise characteristics are easily within requirements. The pixel full well, linearity, read noise, dynamic range, SNR etc. were verified independently against the published datasheet by the vendor. Optical modifications to the current design show that the stock lenses with minimal mechanical fabrication can meet the image adaptation requirements. Thermal stability of the camera was verified in sustained running operation and found to reach equilibrium in about 90 minutes of start and remains within couple of degrees stable thereafter. A force air cooling and extra heat sinks may be required to bring the operating temperature value slightly lower than the current equilibrium value of 47 deg C. In summary, it is feasible to incorporate the Emergent Vision Technologies (EVT) camera model HS-1800-S based on IMX425 monochrome CMOS sensor in GONG system and maintain current data format and SNR characteristics.